\documentclass[prd,nofootinbib,a4paper]{revtex4}
\usepackage[a4paper, hdivide={3cm,,3cm}, vdivide={3cm,,3cm}]{geometry}

\usepackage{amstext,amsfonts,amssymb,amsmath}
\usepackage[english]{babel}
\usepackage[ansinew]{inputenc}
\usepackage{units}
\usepackage{hyperref}

\hypersetup{
    pdftitle={Kinetic and mass mixing with three abelian groups},
    pdfauthor={Julian Heeck, Werner Rodejohann}
}

\begin{document}

\let \Lold \L
\def \L {\mathcal{L}} 
\newcommand{\matrixx}[1]{\begin{pmatrix} #1 \end{pmatrix}} 
\newcommand{\sm}{\mathrm{SM}}

\title{Kinetic and mass mixing with three abelian groups}

\author{Julian \surname{Heeck}}
\email{julian.heeck@mpi-hd.mpg.de}
\affiliation{Max--Planck--Institut f\"ur Kernphysik,\\Postfach 103980, D--69029 Heidelberg, Germany}

\author{Werner \surname{Rodejohann}}
\email{werner.rodejohann@mpi-hd.mpg.de}
\affiliation{Max--Planck--Institut f\"ur Kernphysik,\\Postfach 103980, D--69029 Heidelberg, Germany}


\begin{abstract}
We present the possible mixing effects associated with the low-energy
limit of a Standard-Model extension by two abelian gauge groups
$U(1)_1 \times U(1)_2$. We derive general formulae and approximate
expressions that connect the gauge eigenstates to the mass
eigenstates. Applications using the well-studied groups $U(1)_B$, 
$U(1)_{B-L}$, $U(1)_{L_\alpha - L_\beta}$ ($L_\alpha$ being lepton
flavor numbers), and $U(1)_\mathrm{DM}$ (a symmetry acting only on
the dark matter sector) are discussed briefly.
\end{abstract}

\maketitle


\section{Introduction}

Augmenting the Standard Model (SM) gauge group $G_\sm \equiv SU(3)_C \times SU(2)_L \times U(1)_Y$ by an additional abelian group $U(1)'$ is well motivated by grand unified theories (GUTs)~\cite{Langacker:2008yv}, flavor symmetries~\cite{gaugedflavor}, and dark matter (DM) models~\cite{gaugeddarkmatter}. 
Depending on the symmetry breaking scheme a non-diagonal mass matrix for the neutral gauge bosons is possible, so the physical mass eigenstates are linear combinations of the original gauge eigenstates (henceforth referred to as mass mixing). The precise measurements of the masses and couplings of the SM gauge bosons $Z_\sm$ and $W^\pm$ at LEP put stringent constraints on the mixing parameters and consequently on the symmetry breaking sector.
An entirely different type of mixing is associated with the kinetic terms of the gauge fields:
Since the field strength tensor $F^{\mu\nu}$ of an abelian gauge group
is a gauge invariant object of mass dimension $2$, a renormalizable
Lagrangian can contain non-canonical kinetic cross-terms $\propto
\sin\chi F_1^{\mu\nu} F_{2, \mu\nu}$ if the gauge group includes
$U(1)_1 \times U(1)_2$. The kinetic mixing angle $\chi$ modifies the
coupling of the corresponding gauge bosons and can therefore lead to
observable effects~\cite{kineticmixing}. The case of two abelian
groups -- one of them being the hypercharge gauge group $U(1)_Y$ -- is
well studied and widely used in model building, but the generalization
to more abelian factors is seldom discussed, even though this
structure naturally occurs in some string theory and GUT
models~\cite{renormalization}. Renormalizability of the theory
requires the gauge group to be free of anomalies, which drastically
limits the allowed additional $U(1)'$ groups, unless additional
fermions are introduced. This condition is of course even more
constraining in gauge extensions with several new abelian factors;
even without tapping into the various GUT-inspired symmetries, 
there are several interesting combinations of
well-studied symmetries that lead to valid models, e.g.~$U(1)_L \times U(1)_B$~\cite{pavel}, $U(1)_B\times U(1)_\mathrm{DM}$, or $U(1)_{B-L}\times U(1)_{L_\mu-L_\tau}$.

We will present the generalization of the well-studied gauge group
$G_\sm \times U(1)'$ to $G_\sm \times U(1)' \times U(1)''$, which
introduces three kinetic mixing angles and three mass-mixing
parameters. To demonstrate possible applications in model building we
show that $U(1)_B\times U(1)_\mathrm{DM}$ generates isospin-dependent
nucleon-DM scattering and that $U(1)_{B-L}\times U(1)_{L_\mu-L_\tau}$ can in principle induce non-standard neutrino interactions (NSIs). 

 The remaining part of this work is organized as follows. In Sec.~\ref{sec:kineticandmassmixing} we will derive the connection between gauge and mass eigenstates for the neutral vector bosons and give approximate expressions for the mixing matrix and mass shifts. Specific models for dark matter model building and flavor symmetries will be presented in Sec.~\ref{sec:applications}. We summarize and conclude our findings in Sec.~\ref{sec:conclusion}.


\section{Kinetic and Mass Mixing}
\label{sec:kineticandmassmixing}

The most general effective Lagrange density after breaking $G_\sm \times U(1)_1\times U(1)_2$ to $SU(3)_C\times U(1)_\mathrm{EM}$ can be written as $\L = \L_\sm + \L_{X_1} + \L_{X_2} + \L_\mathrm{mix}$, with
\begin{align}
\begin{split}
	\L_\mathrm{SM} &= -\frac{1}{4} \hat{B}_{\mu\nu} \hat{B}^{\mu\nu} -\frac{1}{4} \hat{W}^a_{\mu\nu} \hat{W}^{a\mu\nu} + \frac{1}{2} \hat{M}_Z^2 \hat{Z}_\mu \hat{Z}^\mu  - \frac{\hat{e}}{\hat{c}_W} j_Y^\mu \hat{B}_\mu -\frac{\hat{e}}{\hat{s}_W} j_W^{a\mu} \hat{W}^a_\mu\,,	\\
	\L_{X_i} &= -\frac{1}{4} \hat{X}_{i\, \mu\nu} \hat{X}_i^{\mu\nu}+ \frac{1}{2} \hat{M}_{X_i}^2 \hat{X}_{i\, \mu} \hat{X}_i^\mu - \hat{g}_i  j_i^\mu \hat{X}_{i\, \mu} \,,\qquad i=1,2\,,\\
	\L_\mathrm{mix} &=-\frac{\sin\alpha}{2} \hat{B}_{\mu\nu} \hat{X}_1^{\mu\nu} -\frac{\sin\beta}{2} \hat{B}_{\mu\nu} \hat{X}_2^{\mu\nu} -\frac{\sin\gamma}{2} \hat{X}_{1\,\mu\nu} \hat{X}_2^{\mu\nu}\\
	&\quad + m_1^2\, \hat{Z}_\mu \hat{X}_1^\mu+ m_2^2\,  \hat{Z}_\mu \hat{X}_2^\mu+ m_3^2\,  \hat{X}_{1\,\mu} \hat{X}_2^\mu\,.
\end{split}
\label{eq:Lagrangian}
\end{align}
The currents are defined as
\begin{align}
\begin{split}
	j_Y^\mu &= -\sum_{\ell = e,\mu,\tau} \left[\overline{L}_\ell \gamma^\mu L_\ell + 2\, \overline{\ell}_R \gamma^\mu \ell_R \right] + \frac{1}{3} \, \sum_{\mathrm{quarks}} \left[\overline{Q}_L \gamma^\mu Q_L + 4 \,\overline{u}_R \gamma^\mu u_R -2\, \overline{d}_R \gamma^\mu d_R\right]\,,\\
	 j_W^{a\mu} &= \sum_{\ell = e,\mu,\tau}\overline{L}_\ell \gamma^\mu \frac{\sigma^a}{2} L_\ell + \sum_{\mathrm{quarks}}  \overline{Q}_L \gamma^\mu \frac{\sigma^a}{2} Q_L \,,
\end{split}
\end{align}
with the left-handed $SU(2)_L$ doublets $Q_L$ and $L_\ell$ and the
Pauli matrices $\sigma^a$. We also define the electromagnetic current
$j_\mathrm{EM} \equiv j_W^{3} + \frac{1}{2}\, j_Y$ and the weak
neutral current $j_\mathrm{NC} \equiv 2 j_W^3 - 2 \hat{s}_W^2
j_\mathrm{EM}$; the currents $j_1$ and $j_2$ are left unspecified for
now. We furthermore define the fields $\hat{A} \equiv \hat{c}_W
\hat{B} + \hat{s}_W \hat{W}_3$ and $\hat{Z} \equiv \hat{c}_W \hat{W}_3
- \hat{s}_W \hat{B}$, corresponding to the photon and the $Z_\sm$
boson in the absence of $\L_\mathrm{mix}$. Here and in the following
we will often omit the Lorentz indices on currents and gauge fields, expressions
such as $j A$ are to be read as $j^\mu A_\mu$.

Due to our parameterization of the kinetic mixing angles, the hypercharge field strength tensor $\hat{B}_{\mu\nu}$ and the field strength tensors $\hat{X}_i^{\mu\nu}$ of $U(1)_1 \times U(1)_2$ share the symmetric mixing matrix
\begin{align}
	\L\supset\, -\frac{1}{4}\,\matrixx{\hat{B}^{\mu\nu}, &
\hat{X}_1^{\mu\nu}, & \hat{X}_2^{\mu\nu}} \matrixx{1 & \sin \alpha &
\sin\beta\\ \cdot & 1 & \sin\gamma\\\cdot & \cdot & 1}
\matrixx{\hat{B}_{\mu\nu}\\ \hat{X}_{1\, \mu\nu}\\ \hat{X}_{2\,
\mu\nu}} .
\end{align}
In complete analogy to Ref.~\cite{kineticmixing} we can transform the gauge fields $(\hat{B}, \hat{X}_1, \hat{X}_2)$ into a basis $(B,X_1,X_2)$ with canonical (diagonal) kinetic terms
\begin{align}
	\matrixx{\hat{B}\\ \hat{X}_1\\ \hat{X}_2} = \matrixx{1 & -
t_\alpha &  (t_\alpha s_\gamma - s_\beta/c_\alpha)/D\\ 0 & 1/c_\alpha
& (t_\alpha s_\beta - s_\gamma/c_\alpha)/D \\ 0 & 0 &
c_\alpha/D}\matrixx{B\\ X_1\\ X_2} ,
	\label{eq:good_kinetic_terms}
\end{align}
where $D\equiv \sqrt{1- s_\alpha^2 - s_\beta^2 - s_\gamma^2 + 2 s_\alpha s_\beta s_\gamma}$, $s_x \equiv \sin x$, $c_x \equiv \cos x$, and $t_x \equiv \tan x$.
The transformation~\eqref{eq:good_kinetic_terms} diagonalizes the kinetic terms and yields the massless photon $A$ and the mass matrix for the massive neutral fields in the basis $(Z,X_1,X_2)$
\begin{align}
	\mathcal{M}^2 = 
	\matrixx{\hat{M}_Z^2 & m_1^2/c_\alpha + \hat{M}_Z^2 \hat{s}_W t_\alpha & M_{13}^2\\
	\cdot & \hat{M}_{X_1}^2/c_\alpha^2 + \hat{s}_W t_\alpha (2 m_1^2 + \hat{M}_Z^2 \hat{s}_W s_\alpha)/c_\alpha &
	M_{23}^2\\
	\cdot & \cdot & M_{33}^2} ,
	\label{eq:massmatrix}
\end{align}
with the three extra long expressions
\begin{align}
\begin{split}
	M_{13}^2 \cdot c_\alpha D &\equiv (\hat{M}_Z^2  \hat{s}_W (s_\beta - s_\alpha s_\gamma) + m_1^2 (s_\alpha s_\beta - s_\gamma) + m_2^2 c_\alpha^2)\,,\\
	M_{23}^2\cdot c_\alpha^2 D &\equiv \hat{M}_{X_1}^2 (s_\alpha s_\beta - s_\gamma) + \hat{M}_Z^2 \hat{s}_W^2 s_\alpha (s_\beta - s_\alpha s_\gamma) +m_1^2 \hat{s}_W (s_\beta - 2 s_\alpha s_\gamma + s_\beta s_\alpha^2) \\
	&\quad + m_2^2 \hat{s}_W s_\alpha c_\alpha^2  + m_3^2 c_\alpha^2 \,,\\
	M_{33}^2\cdot c_\alpha^2 D^2 &\equiv \hat{M}_{X_2}^2 c_\alpha^4 + \hat{M}_{X_1}^2 (s_\gamma - s_\alpha s_\beta)^2 + \hat{M}_Z^2 \hat{s}_W^2 (s_\beta - s_\alpha s_\gamma)^2 \\
	&\quad 	- 2 m_1^2 \hat{s}_W (s_\alpha s_\beta - s_\gamma) (s_\alpha s_\gamma - s_\beta) 
+ 2 m_2^2 c_\alpha^2 \hat{s}_W (s_\beta - s_\alpha s_\gamma)\\
	&\quad+ 2 m_3^2 c_\alpha^2 (s_\alpha s_\beta - s_\gamma)\,.
\end{split}
\end{align}
$\mathcal{M}^2$ is a real symmetric matrix and can therefore be
diagonalized by an orthogonal matrix $U$: $U^T \mathcal{M}^2 U =
\mathrm{diag}(M_1^2, M_2^2, M_3^2)$, with $M_i^2$ being the physical fields. 
This diagonalization introduces
in general three more mixing angles $\xi_i$ that are connected to the
entries in $\mathcal{M}^2$.  
The gauge eigenstates $\hat{A}$, $\hat{Z}$, $\hat{X}_1$, and $\hat{X}_2$ couple to the currents $\hat{e} j_\mathrm{EM}$, $\hat{g}_Z  j_\mathrm{NC}$,\footnote{Here we defined the coupling strength of the $\hat Z$ boson $\hat{g}_Z \equiv \hat{e}/2 \hat{c}_W \hat{s}_W$.} $\hat g_1 j_1$, and $\hat g_2 j_2$, respectively, and are connected to the physical mass eigenstates $A$, $Z_1$, $Z_2$, and $Z_3$ via
\begin{align}
	\matrixx{\hat{A}\\ \hat{Z}\\ \hat{X}_1 \\ \hat{X}_2} = \matrixx{
	1 & 0 & - \hat{c}_W t_\alpha & \hat{c}_W (s_\alpha s_\gamma - s_\beta)/c_\alpha D\\
	0 & 1 & \hat{s}_W t_\alpha & \hat{s}_W (s_\beta - s_\alpha s_\gamma)/c_\alpha D\\
	0 & 0 & 1/c_\alpha & (s_\alpha s_\beta - s_\gamma)/c_\alpha D\\
	0 & 0 & 0 & c_\alpha/D}\matrixx{1 & 0 & 0 & 0\\ 0 & & & \\ 0 & & U &\\ 0 & & & }\matrixx{A\\ Z_1 \\ Z_2 \\ Z_3} ,
	\label{eq:massvsgauge}
\end{align}
or, inverted:
\begin{align}
	\matrixx{A\\ Z_1 \\ Z_2 \\ Z_3}= \matrixx{1 & 0 & 0 & 0\\ 0 & & & \\ 0 & & U^T &\\ 0 & & & } \matrixx{1 & 0 & \hat{c}_W s_\alpha & \hat{c}_W s_\beta\\
	0 & 1 & - \hat{s}_W s_\alpha & -\hat{s}_W s_\beta\\
	0 & 0 & c_\alpha & (s_\gamma - s_\alpha s_\beta)/c_\alpha\\
	0 & 0 & 0 & D/c_\alpha} \matrixx{\hat{A}\\ \hat{Z}\\ \hat{X}_1 \\ \hat{X}_2} .
\end{align}
Due to our parameterization, we can identify $\hat e =e = \sqrt{4\pi \alpha_\mathrm{EM}}$ with the usual electric charge. The physical Weinberg angle is defined via
\begin{align}
	s_W^2 c_W^2 = \frac{\pi \alpha_\mathrm{EM} (M_1)}{\sqrt{2} G_F M_1^2}\,,
\end{align}
which leads to the identity $s_W c_W M_1 = \hat s_W \hat c_W \hat M_Z$~\cite{kineticmixing}.

The general case is complicated to discuss and hardly illuminating, which is why we will work with several approximations from here on out.
In the limit $m_i^2 \ll \hat{M}_Z^2, \hat{M}_{X_j}^2$, $\alpha,\beta,\gamma \ll 1$ the mass matrix~\eqref{eq:massmatrix} simplifies to
\begin{align}
	\mathcal{M}^2 \simeq \matrixx{\hat{M}_Z^2 & \hat{M}_Z^2
\hat{s}_W \alpha + m_1^2 & \hat{M}_Z^2 \hat{s}_W \beta + m_2^2\\ \cdot
& \hat{M}_{X_1}^2 & -\hat{M}_{X_1}^2 \gamma + m_3^2 \\ \cdot & \cdot &
\hat{M}_{X_2}^2} .
\end{align}
Diagonalization leads to the resulting connection between gauge and mass eigenstates
\begin{align}
	\matrixx{\hat{A}\\ \hat{Z}\\ \hat{X}_1 \\ \hat{X}_2} \simeq 
	\matrixx{
	1 & 0 & -\hat{c}_W \alpha & -\hat{c}_W \beta\\ 
	0 & 1 &  \frac{\hat s_W \alpha \hat M_{X_1}^2 + m_1^2}{\hat M_{X_1}^2 - \hat M_Z^2}& \frac{\hat s_W \beta \hat M_{X_2}^2 + m_2^2}{\hat M_{X_2}^2 - \hat M_Z^2}\\ 
	0 & -\frac{\hat s_W \alpha \hat M_{Z}^2 + m_1^2}{\hat M_{X_1}^2 - \hat M_Z^2} & 1 & - \frac{ \gamma \hat M_{X_2}^2 - m_3^2}{\hat M_{X_2}^2 - \hat M_{X_1}^2}\\ 
	0 & - \frac{\hat s_W \beta \hat M_{Z}^2 + m_2^2}{\hat M_{X_2}^2 - \hat M_Z^2} & \frac{ \gamma \hat M_{X_1}^2 - m_3^2}{\hat M_{X_2}^2 - \hat M_{X_1}^2} & 1}\matrixx{A\\ Z_1 \\ Z_2 \\ Z_3} ,
	\label{eq:gaugevsmassapprox}
\end{align}
and one can calculate the mass shift of the $Z$ boson
\begin{align}
	M_{1}^2 / \hat{M}_Z^2 \simeq 1 + \frac{\left(\hat{s}_W \alpha + m_1^2/\hat M_Z^2\right)^2}{1- \hat M_{X_1}^2/\hat{M}_Z^2}+ \frac{\left(\hat{s}_W \beta + m_2^2/\hat M_Z^2\right)^2}{1- M_{X_2}^2/\hat{M}_Z^2}\,.
\end{align}
With this formula we can express $\hat M_Z^2$ in terms of measurable masses:
\begin{align}
	\frac{\hat M_Z^2}{M_1^2} = \frac{s_W^2 c_W^2}{\hat s_W^2 \hat c_W^2} \simeq   1 -\frac{\left({s}_W \alpha + m_1^2/M_1^2\right)^2}{1- M_{2}^2/M_1^2} -\frac{\left({s}_W \beta + m_2^2/M_1^2\right)^2}{1- M_{3}^2 / M_1^2} \,.
\end{align}
The direction of the shift depends on the hierarchy of $\hat M_Z^2$ and $\hat M_{X_i}^2$; a cancellation is possible for $\hat M_{X_1}^2 < \hat{M}_Z^2 < \hat M_{X_2}^2$, which would reduce stringent constraints from the $\rho$ parameter (hiding one $Z'$ with another).
A different way of relaxing the limits on a $Z'$ model by adding additional heavy bosons with specific charges was recently discussed in Ref.~\cite{delAguila:2011yd}.
For completeness we show the effects of heavy $Z'$ bosons in terms of the oblique parameters $S$ and $T$, which can be read off the modified $Z_1$ couplings to $j_W^3$ and $j_\mathrm{EM}$ in the limit $\hat g_{1,2} \equiv 0$~\cite{kineticmixing}:
\begin{align}
\begin{split}
	\alpha_\mathrm{EM} T &\simeq \frac{s_W^2 \alpha^2 - m_1^4/M_1^4}{1-M_2^2/M_1^2}+\frac{s_W^2 \beta^2 - m_2^4/M_1^4}{1-M_3^2/M_1^2}\,,\\
	\alpha_\mathrm{EM} S &\simeq 4 s_W c_W^2 \alpha\, \frac{s_W \alpha + m_1^2/M_1^2}{1-M_2^2/M_1^2} +  4 s_W c_W^2 \beta\, \frac{s_W \beta + m_2^2/M_1^2}{1-M_3^2/M_1^2}\,.
\end{split}
\end{align}


\section{Applications}
\label{sec:applications}

We will now show some applications of the framework laid out above. It
is not our intention to examine the models in complete detail, but
only to consider a few interesting effects. In most cases it suffices
to work with the approximation in Eq.~\eqref{eq:gaugevsmassapprox},
which is used to read off the couplings of the mass eigenstates to the
different currents/particles. Once a proper model is defined by
additional scalars and fermions, one can perform more sophisticated
analyses which make use of numerical diagonalization of the neutral
boson mass matrix in Eq.~\eqref{eq:massmatrix}. In particular, in
specific models the loop-induced kinetic mixing angles can be
calculated.

\subsection{Crossing the streams}
\label{sec:crossterms}

Model building with mixing between $U(1)_1$ and $U(1)_2$ often makes use of the induced coupling of currents, i.e.~$\L_\mathrm{mix}\sim \varepsilon \, j_1 j_2$, which connects the two gauge sectors even if no particle is charged under both groups. We will now derive a necessary condition for such a non-diagonal term at tree level. Taking all of the mixing parameters in Eq.~\eqref{eq:Lagrangian} to be zero except for $m_3$ and $\gamma$, we obtain the coupling of the mass eigenstates $Z_2$ and $Z_3$ to the currents
\begin{align}
	\L \supset \, - \matrixx{\hat g_1 j_1\,, &\hat  g_2 j_2} \matrixx{1 & -t_\gamma\\ 0 & 1/c_\gamma} \matrixx{c_\xi & -s_\xi \\ s_\xi & c_\xi} \matrixx{Z_2\\ Z_3}  \equiv - \matrixx{\hat g_1 j_1\,, & \hat g_2 j_2} V_\gamma U_\xi \matrixx{Z_2\\ Z_3} ,
\end{align}
where $U_\xi$ diagonalizes the mass matrix. Integrating out the heavy mass eigenstates yields an effective four-fermion interaction of the form
\begin{align}
\begin{split}
	\L_\mathrm{eff} &= -\frac{1}{2} \matrixx{\hat g_1 j_1\,, & \hat g_2 j_2} V_\gamma U_\xi \matrixx{1/M_2^2  & 0\\ 0 & 1/M_3^2} U_\xi^T V_\gamma^T \matrixx{\hat g_1 j_1  \\ \hat g_2 j_2}\\
	&= -\frac{1}{2} \matrixx{\hat g_1 j_1\,, & \hat g_2 j_2}
\matrixx{\hat M_{X_1}^2  & m_3^2 \\ m_3^2 & \hat M_{X_2}^2}^{-1}
\matrixx{\hat g_1 j_1  \\ \hat g_2 j_2} .
\end{split}
	\label{eq:fourfermioninteractions}
\end{align}
It is obvious that the coupling matrix is diagonal if $m_3 = 0$, independent of $\gamma$. An analogous calculation can be performed for the coupling of $j_i$ to $j_\mathrm{NC}$ via $m_i$ and $\alpha$, $\beta$, respectively, although it is a bit more tedious because of the additional Weinberg rotation. Nevertheless, the result is the same: an off-diagonal effective coupling $j_i\, j_\mathrm{NC}$ only arises for $m_i\neq 0$, i.e.~$\L_\mathrm{eff} \propto m_{1,2}^2\, j_{1,2} \,j_\mathrm{NC}$.
Since the Weinberg rotation induces a coupling of $j_i$ to the electromagnetic current (first row in Eq.~\eqref{eq:gaugevsmassapprox}), interesting couplings can arise even for $m_{1,2}=0$.

Up until now we discussed only one non-zero $m_i$ and kinetic mixing angle at a time, corresponding to the well-known case of $Z$--$Z'$ mixing. A more general analysis including all our mixing parameters from Eq.~\eqref{eq:Lagrangian} yields the effective four-fermion interactions
\begin{align}
 \L_\mathrm{eff} = -\frac{1}{2} \matrixx{\hat{g}_Z j_\mathrm{NC} \\
\hat{g}_1 j_1 -e \hat{c}_W s_\alpha j_\mathrm{EM} \\ \hat{g}_2 j_2 -e
\hat{c}_W s_\beta j_\mathrm{EM} }^T \matrixx{\hat{M}_Z^2 & m_1^2 &
m_2^2 \\ \cdot & \hat{M}_{X_1}^2 & m_3^2 \\ \cdot & \cdot &
\hat{M}_{X_2}^2}^{-1}  \matrixx{\hat{g}_Z j_\mathrm{NC} \\ \hat{g}_1
j_1 -e \hat{c}_W s_\alpha j_\mathrm{EM} \\ \hat{g}_2 j_2 -e \hat{c}_W
s_\beta j_\mathrm{EM} } .
\label{eq:threeflavormixing}
\end{align}
Because the $3\times 3$ coupling matrix takes the explicit form
\begin{align}
 \matrixx{\hat{M}_Z^2 & m_1^2 & m_2^2 \\ \cdot & \hat{M}_{X_1}^2 & m_3^2 \\ \cdot & \cdot & \hat{M}_{X_2}^2}^{-1} = 
\frac{1}{\Delta^6}
\matrixx{\hat{M}_{X_1}^2 \hat{M}_{X_2}^2 - m_3^4 & -m_1^2
\hat{M}_{X_2}^2 + m_2^2 m_3^2 & - \hat{M}_{X_1}^2 m_2^2 + m_1^2 m_3^2
\\ \cdot & \hat{M}_{Z}^2 \hat{M}_{X_2}^2 - m_2^4 & - \hat{M}_{Z}^2
m_3^2 + m_1^2 m_2^2 \\ \cdot & \cdot & \hat{M}_{Z}^2 \hat{M}_{X_1}^2 -
m_1^4} ,
\end{align}
with $\Delta^6 \equiv \hat{M}_Z^2 \hat{M}_{X_1}^2 \hat{M}_{X_2}^2 - \hat{M}_Z^2 m_3^4 - \hat{M}_{X_2}^2 m_2^4 - \hat{M}_{X_2}^2 m_1^4 + 2 m_1^2 m_2^2 m_3^2$, we end up with new off-diagonal couplings like $m_2^2 m_3^2 \, j_1 \, j_\mathrm{NC}$, even if there is no direct coupling $m_1^2 \, j_1 \, j_\mathrm{NC}$.

\subsection{\texorpdfstring{$\boldsymbol{U(1)_B\times U(1)_\mathrm{DM}}$}{U(1)(B) x U(1)(DM)}}

It was recently shown that the seemingly incompatible results of the
dark matter (DM) direct detection experiments DAMA/CoGeNT and XENON can be alleviated with the introduction of isospin-dependent couplings of nucleons to dark matter~\cite{darkmatteranomaly}. One of the models used in Ref.~\cite{Frandsen:2011cg} to explain this coupling is based on gauged baryon number $U(1)_1 \equiv U(1)_B$.\footnote{It was pointed out in Ref.~\cite{gaugedbaryonnumber} that a gauge boson coupled to the baryon number $B$ can be light. The drawback of such a symmetry is the unavoidable introduction of new chiral fermions to cancel occurring triangle anomalies. An anomaly-free symmetry (SM + right-handed neutrinos) with similarly weak constraints is $U(1)_{B-3 L_\tau}$~\cite{B-3Ltau}.}
With dark matter charged under this gauge group, the resulting cross
section turns out to be too small to explain the observed events,
unless the coupling of $Z'$ to dark matter is significantly stronger
than to quarks (i.e.~DM carries a large baryon number). However, in a model with another gauge group $U(1)_2 \equiv U(1)_\mathrm{DM}$ -- acting only on the DM sector~\cite{gaugeddarkmatter} -- the dark matter coupling constant $g_\mathrm{DM}$ can be naturally large compared to $g_B$, which allows for a sizable cross section as long as the mass mixing between the groups is not too small.\footnote{A similar model was proposed very recently in the same context, see Ref.~\cite{erster}.}
We only introduce one DM Dirac fermion $\chi$, so the $U(1)_2$ current takes the form $j^\mu_2 = j^\mu_\mathrm{DM} = \overline{\chi} \gamma^\mu \chi$. For clarity we take all mixing parameters in Eq.~\eqref{eq:Lagrangian} to be zero -- except for $m_3$ and $\beta$ -- and assume $Z_{2,3}$ to be light ($M_{2,3}^2 \ll  M_1^2$) to generate a large cross section. Eq.~\eqref{eq:gaugevsmassapprox} then gives the approximate couplings
\begin{align}
\begin{split}
	\L \supset \ &-\left( \frac{e}{2 c_W s_W} j_\mathrm{NC} + \beta s_W g_\mathrm{DM} j_\mathrm{DM}\right) Z_1
	- \left( g_B j_B - g_\mathrm{DM} \frac{m_3^2}{M_3^2 - M_2^2} j_\mathrm{DM}\right) Z_2\\
	&-\left(g_\mathrm{DM} j_\mathrm{DM} - \beta c_W e j_\mathrm{EM} +g_B \frac{m_3^2}{M_3^2-M_2^2} j_{B}\right) Z_3\,.
\end{split}
\end{align}
These terms couple dark matter to nucleons via $m_3$, and because of $\beta$, proton and neutron couple differently, i.e.~the interaction is isospin-dependent. Integrating out all the gauge bosons gives the effective vector-vector interactions in the usual parameterization
\begin{align}
	\L_\mathrm{eff} \supset \, f_p\, \overline{\chi} \gamma_\mu \chi \,\overline{p}\gamma^\mu p + f_n\, \overline{\chi} \gamma_\mu \chi \,\overline{n}\gamma^\mu n\,,
\end{align}
with the ratio of the neutron and proton couplings
\begin{align}
	f_n/f_p = \frac{1}{1+r}\,, &&
	r \simeq e c_W \frac{\beta}{g_B} \frac{M_2^2}{m_3^2}\,.
	\label{eq:fnfp}
\end{align}
We can easily find parameters to generate $f_n/f_p\simeq -0.7$ ($r\simeq -2.4$). The overall DM-neutron cross section can be calculated to be~\cite{Jungman:1995df}
\begin{align}
\begin{split}
	\sigma_n &= \frac{1}{64\pi}\left( \frac{m_\chi m_n}{m_\chi + m_n}\right)^2 f_n^2
	\simeq \frac{m_n^2}{64\pi}\, \left(g_B g_\mathrm{DM} \frac{m_3^2}{M_2^2 M_3^2}\right)^2
	\simeq  2\, \alpha_\mathrm{DM} \beta^2 \left(\frac{\unit[1]{GeV}}{M_{3}}\right)^4 \, \unit[10^{-31}]{cm^2}\,,
\end{split}
\end{align}
where we defined $\alpha_\mathrm{DM} \equiv g^2_\mathrm{DM}/4\pi$ and
assumed $m_\chi \gg m_n$. To obtain the last equation we replaced
$g_B\, m_3^2$ with the demanded value for $r$ from
Eq.~\eqref{eq:fnfp}. For $\beta \sim 10^{-3}$ it is possible to
generate the required DAMA/CoGeNT cross section $\sigma_n \sim
10^{-38}$--$\unit[10^{-37}]{cm^2}$~\cite{darkmatteranomaly} without
being in conflict with other
constraints~\cite{gaugedbaryonnumber,kineticmixinglimits}. We note
that the dark matter fine-structure constant $\alpha_\mathrm{DM}$ is
not restricted to be small. 

Due to the required non-zero $m_3^2$ we will have a non-trivial scalar sector that also serves as a mediator between the SM and the dark sector. We assume these scalars to be heavy enough to not alter our foregoing discussion.

Aside from the group $U(1)_B\times U(1)_\mathrm{DM}$ discussed above, further interesting models using this framework in the dark matter sector could be build using leptophilic groups like $U(1)_{L_\mu-L_\tau}\times U(1)_\mathrm{DM}$, with the possibility to resolve the PAMELA positron excess via the small leptophilic admixture~\cite{Baek:2008nz}.

\subsection{\texorpdfstring{$\boldsymbol{U(1)_{B-L}\times U(1)_{L_\mu-L_\tau}}$}{U(1)(B-L) x U(1)(Lmu-Ltau)}}

A family non-universal model can be build using $U(1)_1 \equiv U(1)_{B-L}$ and $U(1)_2 \equiv U(1)_{L_\mu-L_\tau}$ without introducing anomalies. Each group is anomaly-free if the Standard Model is extended with $3$ right-handed neutrinos $N_{i, R}$ carrying appropriate lepton numbers, so the only potential triangle anomalies involve both gauge groups:
\begin{align}
\begin{split}
	U(1)_{L_\mu-L_\tau}  &- U(1)_{B-L} - U(1)_{L_\mu-L_\tau}: \\
	 &\sum_{\mu,\tau} Y_{B-L} = 2 \left[Y_{B-L} (\mu_L)+Y_{B-L} (\nu_\mu) +Y_{B-L} (\mu^c_R)+Y_{B-L} (N^c_{2, R})\right] = 0\,,\\
	U(1)_{L_\mu-L_\tau}  &- U(1)_{B-L} - U(1)_{B-L}: \\
	& \sum_{\mu,\tau} Y_{L_\mu - L_\tau}\, Y_{B-L}^2 = \sum_{\mu} Y_{B-L}^2 - \sum_{\tau} Y_{B-L}^2 = 0\,,\\
	U(1)_{L_\mu-L_\tau}  &- U(1)_{B-L} - U(1)_Y: \\
	& \sum_{\mu,\tau}  Y_{L_\mu - L_\tau}\, Y_{B-L}\, Y = \sum_{\mu}  Y_{B-L}\, Y - \sum_{\tau} Y_{B-L}\, Y = 0\,,\\	
\end{split}
\end{align}
where the last two relations follow from the universality of $U(1)_Y$
and $U(1)_{B-L}$. The anomalies from $SU(2) - U(1)_1 - U(1)_2$ and
$SU(3) - U(1)_1 - U(1)_2$ vanish trivially in any model due to the
tracelessness of the non-abelian generators. The same conclusion can,
of course, be reached for any of the anomaly-free $L_\alpha - L_\beta$
symmetries. However, $L_\mu - L_\tau$ is favored over $L_e - L_\mu$
and $L_e - L_\tau$ because of a more reasonable flavor structure of the
neutrino mass matrix~\cite{wir_alt}. 

The gauge boson 
$Z_2 = Z_{B-L}$ is highly constrained by collider experiments
($M_{B-L}/g_{B-L} \gtrsim 6$--$\unit[7]{TeV}$ at
$95\%$~C.L.~\cite{LEP-2bounds}),\footnote{The limits from LEP~2 and
Tevatron have been derived under the assumption of just one additional
gauge boson, but still hold approximately when additional bosons are
included~\cite{delAguila:2011yd}.} but $Z_3 = Z_{L_\mu - L_\tau}$ can
have a mass around the electroweak scale and there is actually a
preferred region around $M_{L_\mu-L_\tau}/g_{L_\mu-L_\tau} \simeq
\unit[200]{GeV}$ that ameliorates the tension between the theoretical
and experimental values for the anomalous magnetic moment of the
muon~\cite{wir} (see~\cite{Lmu-Ltau} for earlier works).

In $U(1)_{B-L}\times U(1)_{L_\mu-L_\tau}$ models with non-vanishing mass mixing the parameter $m_3$ induces an effective coupling of the currents $j_{L_\mu-L_\tau}$ and $j_{B-L}$ (see Sec.~\ref{sec:crossterms}), which leads for example to non-standard neutrino interactions, usually parameterized by the non-renormalizable effective Lagrangian 
\begin{align}
	\L_\mathrm{eff}^\mathrm{NSI} = -2 \sqrt{2} G_F \varepsilon_{\alpha\beta}^{f P} \left[\bar f \gamma^\mu P f \right] \left[ \bar \nu_\alpha \gamma_\mu P_L \nu_\beta\right] .
\end{align}
The model at hand induces $\varepsilon_{\mu\mu}^{f P} = - \varepsilon_{\tau\tau}^{f P}$, easily read off from Eq.~\eqref{eq:fourfermioninteractions}:
\begin{align}
\begin{split}
	\varepsilon^{e V}_{\mu\mu} &\simeq - \frac{1}{2 \sqrt{2} G_F} g_1
g_2 \frac{m_3^2}{M_{2}^2 M_3^2} \simeq -2\times 10^{-6}\,\frac{1}{g_1
g_2} \left(\frac{m_3}{\unit[10]{GeV}}\right)^2  \left(
\frac{\unit[6]{TeV}}{M_2/g_1}\right)^2  \left(
\frac{\unit[200]{GeV}}{M_3/g_2}\right)^2 ,\\
	\varepsilon^{u V}_{\mu\mu} &= \varepsilon^{d V}_{\mu\mu} = - \varepsilon^{e V}_{\mu\mu}/3\,,
\end{split}
\label{eq:nsis}
\end{align}
which are in general too small to be observable in current experiments~\cite{nsi}. 
Larger NSIs can be generated at the price of introducing mass mixing of $Z_{L_\mu-L_\tau}$ with $Z_\sm$ via $m_2$ (using the more general Eq.~\eqref{eq:threeflavormixing}). Even though this kind of mixing is highly constrained by collider experiments, the arising NSIs are testable in future facilities for $M_2 < M_1$~\cite{wir}.
Substituting $U(1)_{B-L}$ in Eq.~\eqref{eq:nsis} with less
constrained symmetries like $U(1)_B$ or $U(1)_{B-3 L_\tau}$ (including
fermions to cancel arising anomalies) allows for lighter gauge bosons
and therefore also larger NSIs; a recent discussion of additional
constraints on $Z'$ bosons with non-universal couplings to charged leptons can be found in Ref.~\cite{Chiang:2011cv}. 
Since our framework does not involve mixing with the SM
gauge bosons -- at least at tree level -- the bounds on the mixing
parameters are less stringent.


\section{Conclusion}
\label{sec:conclusion}

The extension of the Standard Model by an additional abelian factor
$U(1)'$ is a well motivated and frequently discussed area in model building. It
is not far fetched to extend this even further to $G_\sm \times
[U(1)']^n$, provided the full gauge group stays free of anomalies. We
discussed the most general low-energy Lagrangian for the case $n=2$,
including kinetic mixing among the abelian groups. We showed how the
mixing among several gauge groups -- such as $U(1)_{B-L}$,
$U(1)_{L_\mu-L_\tau}$, and $U(1)_\mathrm{DM}$ -- 
can lead to interesting effects like non-standard neutrino
interactions and isospin-dependent dark matter scattering. This opens
up new and interesting possibilities in model building.

\begin{acknowledgments}
This work was supported by the ERC under the Starting Grant 
MANITOP and by the DFG in the Transregio 27. JH acknowledges support by the IMPRS-PTFS.
\end{acknowledgments}

\end{document}